\documentclass[%
 reprint,
 amsmath,amssymb,superscriptaddress,
 aps,
prl,
]{revtex4-1}

\usepackage{graphicx}
\usepackage{dcolumn}
\usepackage{bm}

\begin{document}

\title{Intertwined Spin and Orbital Density Waves in MnP Uncovered by Resonant Soft X-ray Scattering}

\author{B. Y. Pan}
\affiliation{State Key Laboratory of Surface Physics, Department of Physics, and Advanced Materials
Laboratory, Fudan University, Shanghai 200433, China}
\affiliation{School of Physics and Optoelectronic Engineering, Ludong University, Yantai, Shandong 264025, China}
\author{H. Jang}
\author{J.-S. Lee}
\affiliation{Stanford Synchrotron Radiation Lightsource, SLAC National Accelerator Laboratory,
Menlo Park, CA 94025, USA}
\author{R. Sutarto}
\author{F. He}
\affiliation{Canadian Light Source, Saskatoon, Saskatchewan S7N 2V3, Canada}
\author{J. F. Zeng}
\affiliation{Beijing National Laboratory for Condensed Matter Physics, and Institute of Physics, Chinese Academy of Sciences, Beijing 100190, China}
\affiliation{University of Chinese Academy of Sciences, Beijing 100049, China}
\author{Y. Liu}
\affiliation{Center for Correlated Matter, Zhejiang University, Hangzhou, 310058, China}
\author{X. W. Zhang}
\author{Y. Feng}
\author{Y. Q. Hao}
\affiliation{State Key Laboratory of Surface Physics, Department of Physics, and Advanced Materials
Laboratory, Fudan University, Shanghai 200433, China}
\author{J. Zhao}
\affiliation{State Key Laboratory of Surface Physics, Department of Physics, and Advanced Materials
Laboratory, Fudan University, Shanghai 200433, China}
\affiliation{Collaborative Innovation Centre of Advanced Microstructures, Nanjing 210093, China}
\author{H. C. Xu}
\affiliation{State Key Laboratory of Surface Physics, Department of Physics, and Advanced Materials
Laboratory, Fudan University, Shanghai 200433, China}
\author{Z. H. Chen}
\affiliation{Shanghai Institute of Applied Physics, Chinese Academy of Sciences, Shanghai Synchrotron Radiation Facility, Shanghai 201800, China}
\author{J. P. Hu$^\dag$}
\affiliation{Beijing National Laboratory for Condensed Matter Physics, and Institute of Physics, Chinese Academy of Sciences, Beijing 100190, China}
\affiliation{Kavli Institute of Theoretical Sciences, University of Chinese Academy of Sciences, Beijing, 100190, China}
\affiliation{Collaborative Innovation Center of Quantum Matter, Beijing 100084, China}
\author{D. L. Feng$^\ast$}
\affiliation{State Key Laboratory of Surface Physics, Department of Physics, and Advanced Materials
Laboratory, Fudan University, Shanghai 200433, China}
\affiliation{Collaborative Innovation Center of Advanced Microstructures, Nanjing 210093, China}

\begin{abstract}

Unconventional superconductors are often characterized by numerous competing and even intertwined orders in their phase diagrams. In particular,  the  electronic nematic phases,  which spontaneously break rotational symmetry and often simultaneously involve spin, charge and/or orbital orders,  appear conspicuously in both the cuprate and iron-based superconductors.  The fluctuations associated with these phases may provide the exotic pairing glue that underlies their high-temperature superconductivity. Helimagnet MnP, the first Mn-based superconductor under pressure \cite{Cheng2015, Norman2015, Wang2016}, lacks high rotational symmetry. However our resonant soft X-ray scattering (RSXS) experiment discovers novel helical orbital density wave (ODW) orders in this three-dimensional, low-symmetry system, and reveals  intertwined ordering phenomena in unprecedented detail. In particular, a ODW forms with half the period of the spin order and fully develops slightly above the spin ordering temperature; their domains develop simultaneously, yet the spin order domains are larger than those of the ODW; and they
cooperatively produce another ODW with 1/3 the period of the spin order. These observations provide a comprehensive picture of the intricate interplay between spin and orbital orders in correlated materials, and they suggest that nematic-like physics  ubiquitously
exists beyond two-dimensional and high-symmetry systems, and the superconducting mechanism of MnP is likely analogous to those of cuprate and iron-based  superconductors.

\end{abstract}

\maketitle

In strongly correlated electron systems,
the kinetic and interaction energies of electrons compete, which often couples the  charge, spin, and orbital degrees of freedoms, resulting in  a variety of complex quantum phases.
Nematic order is one of the most important forms of intertwined order, which exists in both the two known families of high temperature superconductors --- cuprates and iron-based superconductors.
In iron pnictides, e.g. LaFeAsO and BaFe$_2$As$_2$\cite{Yang2010,Chu2010,Yi2011}, the nematic orbital order often emerges at a  higher temperature than the collinear antiferromagnetic order, and with half the period, indicating that the nematicity may be strongly tied to magnetic fluctuations.
In the stripe phase of the cuprates, the charge order forms with half the period of the spin order\cite{Tranquada1995,Croft2014}.
Moreover, unlike magnetic order, the nematic order is time-reversal invariant, allowing it to directly couple to orders arising from charge and orbital degrees of freedom.
Furthermore, it also appears that superconductivity is maximized in the  region with the strongest nematic fluctuations. In the iron pnictides, for example, the orbital and/or spin fluctuations associated with the nematic  order are suggested to promote superconductivity \cite{Kontani2010,Kuroki2008}.
All these essential phenomena suggest that the electronic nematic phase stands as a bridge  between high temperature superconductivity and magnetism.

\begin{figure*}[]
\centering
\includegraphics[clip,width=16cm]{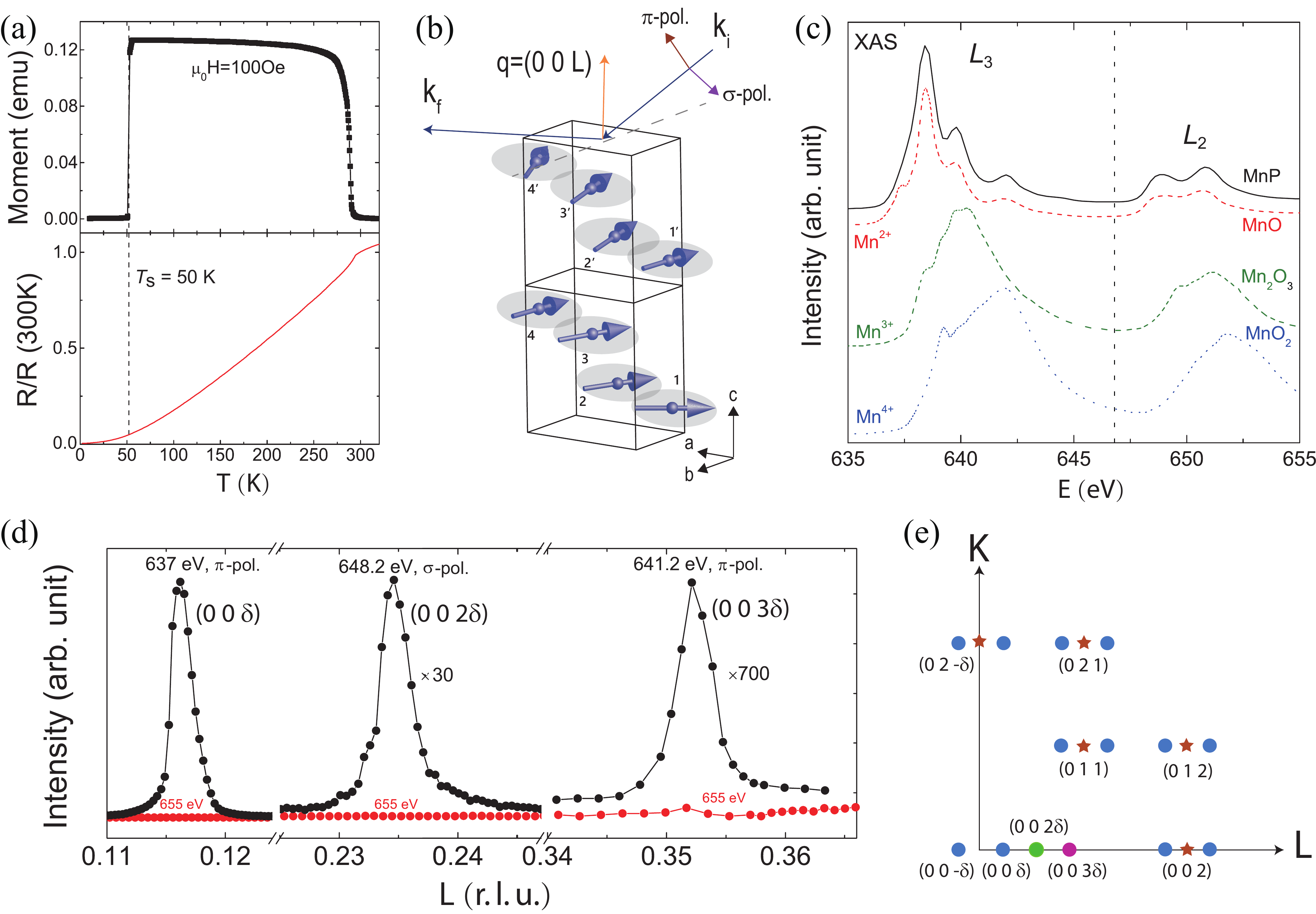}
\centering
\caption{Basic properties and X-ray diffraction peaks of MnP. (a)  Magnetic susceptibility and resistance data of the MnP single crystal sample. (b) Schematic illustration of the helical spin order and the RSXS scattering geometry. The spin rotates about 21$^{\circ}$ between Mn1 and Mn2, and between Mn3 and Mn4, giving an incommensurate double helical spin order, or a helical spin density wave (SDW), with a period of about 8.6$c$. Mn1 and Mn3 spins form one spin helix, and Mn2 and Mn4 spins forms the other.
The incident X-ray is linearly polarized. (c) X-ray absorption spectroscopy (XAS) of MnP (solid black line) and three reference compounds MnO (dashed red line), Mn$_2$O$_3$ (dashed olive line), and MnO$_2$ (dashed blue line) around the Mn $L$-edge. The reference spectra were obtained on beam line 8.0.3 at the Advanced Light Source (ALS) \cite{Qiao2013} and are plotted here offset by 1.55~eV in order to match our MnP spectrum obtained at the SSRL. The Mn $L$-edge XAS of MnP is characteristic of Mn$^{2+}$ ions. (d) Scattering peaks around $q_1$=(0~0~0.116), $q_2$=(0~0~0.234), and $q_3$=(0~0~0.352) at resonant energies (black dots) and non-resonant energy (red dots, E=655~eV) at 20~K. $q_1$ and $q_2$ signals were measured by a Schottky barrier photodiode, and the $q_3$ signals were measured by a quantum-efficient channeltron. The three resonant peaks represent the strongest resonant signals at each q position (see Fig. 3). (e) Schematic diagram of the location of the diffraction peaks in MnP including lattice peaks (red stars), magnetic peaks (blue dots), and the newly discovered resonance peaks in this paper (green and magenta dots).}
\end{figure*}

\begin{figure*}[]
\centering
\includegraphics[clip,width=18cm]{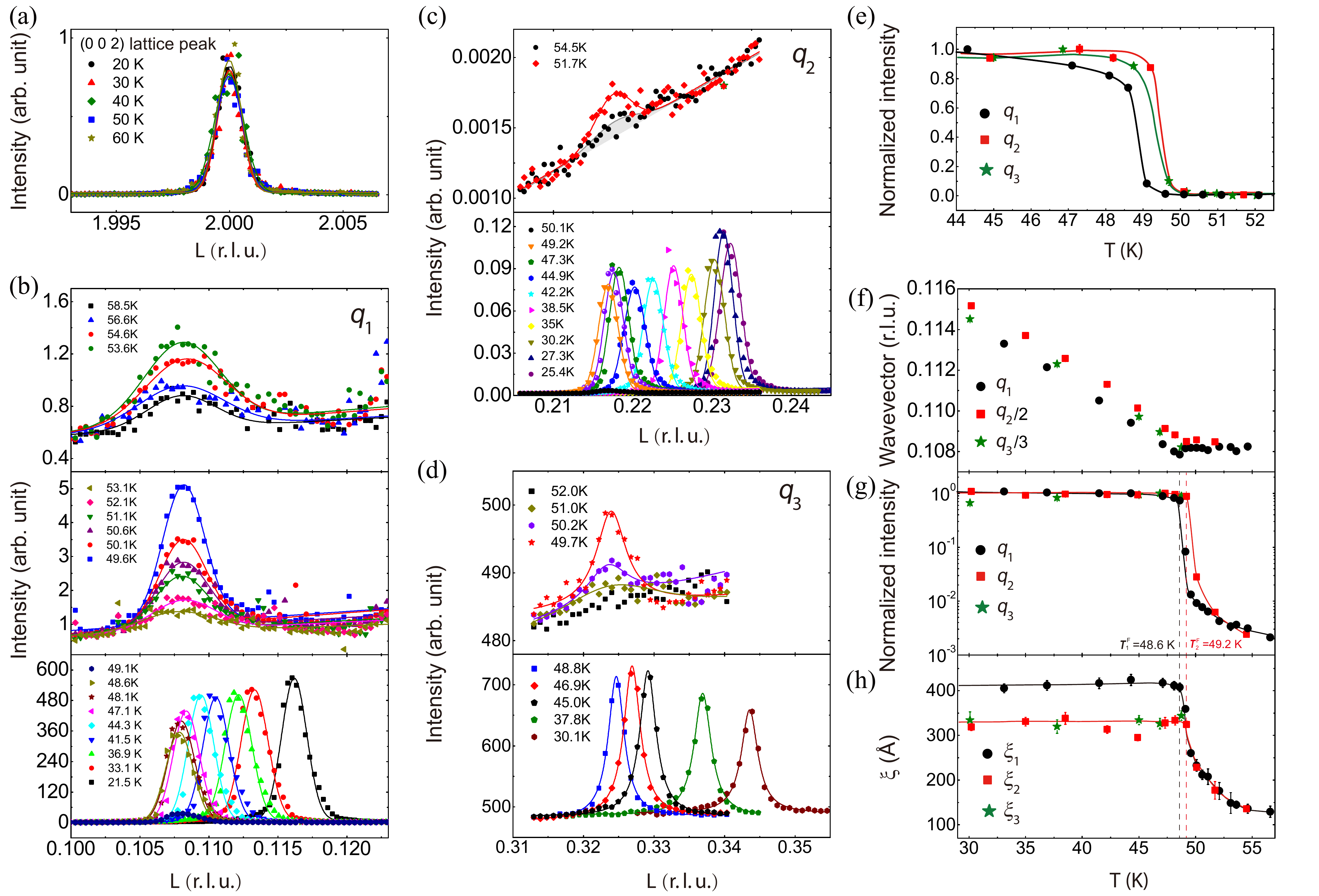}
\caption{Temperature dependences of the diffractions peaks in MnP. (a) (0~0~2) lattice peak measured at 2180~eV from 20 to 60~K. (b-d), $q_1$ $L_3$ (E = 637~eV, $\pi$ polarization), $q_2$ $L_2$ (E = 648.8~eV, $\pi$ polarization), and $q_3$ $L_3$ (E = 636.6~eV, $\pi$ polarization) resonance peaks at different temperatures, measured on cooling. The solid lines in each panel are Gaussian (for $q_1$ and $q_2$ resonance peaks) or Lorentz (for $q_3$ resonance peaks) fittings with linear background. The shadowed area in the upper panel of (c) is to indicate the weak peak intensity at the highest measured temperature. (e) Temperature dependence of the peak area of $q_1$, $q_2$ and $q_3$ near the helical transition, revealing their differing transition temperatures. The data were obtained in the cooling processes. Solid lines are guides to the eye. (f) Temperature dependences of the $q_1$, $q_2$/2, and $q_3$/3. Here, we used a photodiode detector for $q_1$, a channeltron detector for $q_2$, and a Greateyes CCD  for $q_3$. (g) and (h) are the temperature dependences of the peak area and 1/FWHM for $q_1$ (black circles) $q_2$ (red squares) resonance peaks in the cooling measurements, respectively. Log scale were used for the vertical axes and the solid lines are guides to the eye.}
\end{figure*}

Under pressure, the helical magnetic compound CrAs and MnP  have been recently found to be the first Cr- and Mn- based superconductor, respectively \cite{Cheng2015,Wu2014}. For example, under ambient pressure, MnP first enters a ferromagnetic state at $T_C$ = 290~K, then a metamagnetic transition at $T_S$ = 50~K switches it into a double helical magnetic state with moments lying in the $ab$ plane (Fig. 1(a, b)). Such a complex  magnetic behavior suggests that MnP is most likely an unconventional superconductor like the cuprate and iron pnictides. However,
not only has the material a  helical magnetic state that differs from those of cuprates and iron-based superconductors (Fig. 1(b)), but its lattice structure is three dimensional and does not possess any high symmetry rotational axis.
Moreover, the spin, charge and orbital degrees of freedom in MnP could be highly interconnected. Therefore, MnP provides a novel playground to probe the
relation between unconventional superconductivity, magnetism, and possibly nematicity in the vicinity of a helical spin order on a low-symmetry lattice.


Resonant soft X-ray scattering  (RSXS), which can be viewed as a combination of x-ray absorption (XAS) and x-ray emission (XES) spectroscopies with x-ray scattering, provides a direct and powerful probe of the ordering in 3$d$ transition metal compounds\cite{Murakami1998,Peter2004,Peter2005}. To elucidate the electronic state of MnP, we first present its  XAS spectrum at Mn $L$-edge in Fig. 1(c),  together with that of MnO as a fingerprint of the Mn$^{2+}$ valence state\cite{Qiao2013}, Mn$_2$O$_3$ as a typical Mn$^{3+}$ spectrum, and MnO$_2$ representing Mn$^{4+}$. Evidently, MnP has the typical absorption spectrum of a Mn$^{2+}$ state with the half filling high spin configuration. To our knowledge, this is the first XAS measurement on MnP. The Mn$^{2+}$ state in MnP is contrary to the previously speculated Mn$^{3+}$ valence in this material\cite{Goodenough1964}. The spin moment on each Mn site is only $\sim$ 1.3$\mu_B$ \cite{Huber1964}; this value is significantly reduced from the localized spin-only moment $\sim$ 5$\mu_B$ for the 3d$^5$ configuration in Mn$^{2+}$ by Hund's rule. This low spin moment may arise from quantum fluctuations and hybridization\cite{Simonson2012, McNally2014}, as is the case for the iron-based superconductors\cite{Cruz2008,Dai2015}. It should be noted that for a half-filled electronic system, the orbital moment usually should be quenched, thus prohibiting spin-orbital coupling. However, orbital anisotropy/ordering have been found in several half-filled systems due to hybridization anisotropy\cite{Kim2006} and vacancy modulation\cite{Jang2015}. Therefore, orbital angular momentum may be partially restored.

Figure 1(b) shows the scattering geometry, where $bc$ was used as the scattering plane. The incident X-ray is either horizontally ($\pi$) or vertically ($\sigma$) polarized. In this configuration, $\sigma$ polarization has electric field along the $a$ axis and momentum transfer along the (0~0~$L$) direction. When helical magnetic order emerges below $T_S$ = 50~K, its spin moments (blue arrows) lie in the $ab$ plane with magnetic propagation wavevector (0~0~0.117)\cite{Felcher1966}. In order to avoid specular reflection, we used a MnP single crystal with a (101) surface and tilted the sample by  48.4$^{\circ}$ in our RSXS experiments.

We made an exhaustive search for scattering signals at the Mn $L$-edge along (0~0~$L$) at 20~K. By varying $L$, X-ray polarization, photon energy, and detection mode, we discovered resonance peaks at $q_1$=(0~0~0.11634$\pm$0.00002), $q_2$=(0~0~0.23470$\pm$0.00003), and $q_3$=(0~0~0.35239$\pm$0.00004). $q_2$ and $q_3$ wavevectors approximately double and triple that of $q_1$, respectively. Figure 1(d) shows the scans for the strongest resonance peaks at $q_1$ (E = 637~eV, $\pi$ polarization, black dots), $q_2$ (E = 648.2~eV, $\sigma$ polarization, black dots),  $q_3$ (E = 641.2~eV, $\pi$ polarization, black dots), and the corresponding non-resonant scans at 655~eV (red dots). $q_1$ is perfectly consistent with the magnetic diffraction peak q$_m$ = (0~0~$\delta$) in MnP, thus should correspond to helical magnetism. It should be noted that the magnetic diffraction peaks of MnP only appear at [$H$~$K$~$L\pm\delta$] positions in which [$H$~$K$~$L$] represent the Bragg peaks from the lattice, a law universal to helical magnets that has been verified by neutron diffraction in MnP \cite{Yamazaki2010}, FeP\cite{Felcher1971}, and CrAs\cite{Shen2016}. It is thus clear that the new diffraction peaks at $q_2$ and $q_3$ discovered by RSXS are not from magnetism. The diffraction positions of lattice, magnetic, and our newly discovered peaks in reciprocal space are illustrated in Fig.~1(e).

Next, we study the temperature dependences of the diffraction peaks. First, we take the (0~0~2) lattice diffraction peak as a reference. It does not show any observable temperature dependence across $T_S$ and down to 20~K (Fig.~2(a)), indicating no structure transition in this  temperature range. In contrast, $q_1$, $q_2$, and $q_3$ peaks exhibit drastic temperature dependences. Figures 2(b-d) show the three diffraction peaks  taken with their corresponding resonant scattering conditions,
at $q_1$ (E = 637~eV, $\pi$ polarization), $q_2$ (E = 648.8~eV, $\pi$ polarization), and $q_3$ (E = 636.6~eV, $\pi$ polarization) at different temperatures in a cooling sequence, respectively. As can be seen, not only does the  peak intensity rapidly grow across $T_S$, but the propagation wave vector moves to higher $q$ with decreasing temperature. The peak areas of the $q_1$, $q_2$ and $q_3$ diffraction peaks are plotted as a function of temperature in Fig.~2(e), all showing a jump just around 50~K, consistent with a first order transition (the hysteresis behavior of the $q_1$ peak can be found in  Supplementary Fig.~S4(a)).
Interestingly, the $q_1$ peak is fully developed at a slightly lower temperature than the $q_2$ peak, while the full-development temperature of the $q_3$ peak  lies in between or similar to the $q_2$ peak. This observation can be justified by more temperature dependence measurements (Supplementary Fig.~S4(b)).  In  Fig.~2(f), we show the wavevector evolution with temperature, in which $q_2$ and $q_3$ are divided by 2 and 3, respectively, in order to scale with $q_1$. The propagation wavevectors of the three resonance peaks all show pronounced temperature dependence below $T_S$, which is typical for an incommensurate electronic order. Throughout the measured temperature range, $q_2$ and $q_3$ are approximately at the 2$q_1$ and 3$q_1$ positions within the experimental accuracy, respectively, indicating that these electronic orders are interconnected with each other.
According to the upper panels of Fig.~2(b-d), there are detectable  scattering intensities above $T_S$. The $q_1$ and $q_2$ peaks persist up to the highest measured temperatures, i.e., 58.5K and 54.5 K, respectively (upper panels of Fig.~2(b) and Fig.~2(c)). However, $q_3$ peak intensity is not detectable above 52 K (upper panel of Fig.~2(d)) due to its weak intensity and insufficient sensitivity of the detector.  To reveal the detailed evolution at high temperatures, Fig.~2(g) plots the peak areas as a function of temperature in the cooling process in log scale, and
Fig.~2(h) shows the temperature dependences of the  correlation length, $\xi_i$, which is defined as 1/FWHM, reflecting the average domain size along $c$.
Above 50K,  there is a similar slow-growing behavior  for both the $q_1$ and $q_2$ peaks in  Figs.~2(g) and 2(h). The peak intensities are two orders of magnitude lower than their full values, thus they are due to fluctuating orders,
and the long tails into  high temperatures are likely related to local strain distributions. A similar behavior has been observed for the nematic order in iron pnictides under uniaxial strain \cite{Chu2010}.
With  decreasing temperature, $\xi_1$ and $\xi_2$ quickly increase almost identically before they saturate at $T_1^F$ and $T_2^F$, respectively, indicating the peaks are from the same domain and have the same onset temperature, $T^E$, which can be higher than 58.5K, the highest measured temperature.
Below $T_2^F$, $\xi_1$ continues to increase until $T_1^F$, so it is larger than $\xi_2$ at low temperatures.
Meanwhile, the peak intensities show sudden jumps to their fully-developed values just before $T_1^F$ and $T_2^F$ as well.
Therefore, $T_1^F$, $T_2^F$ and $T_3^F$ are defined as the  temperatures that the $q_1$, $q_2$ and $q_3$ peaks are fully developed, respectively.
Note that, between $T_2^F$ and $T_1^F$, the $q_1$ peak intensity is low, indicating that the spin order is still fluctuating while the order corresponding to $q_2$ is already static.
 In addition, we found that $\xi_3$ is almost identical to $\xi_2$.
These behaviors contain rich information on the evolutions of the orders corresponding the diffraction peak, which will be discussed later.

\begin{figure*}[t]
\centering
\includegraphics[clip,width=15cm]{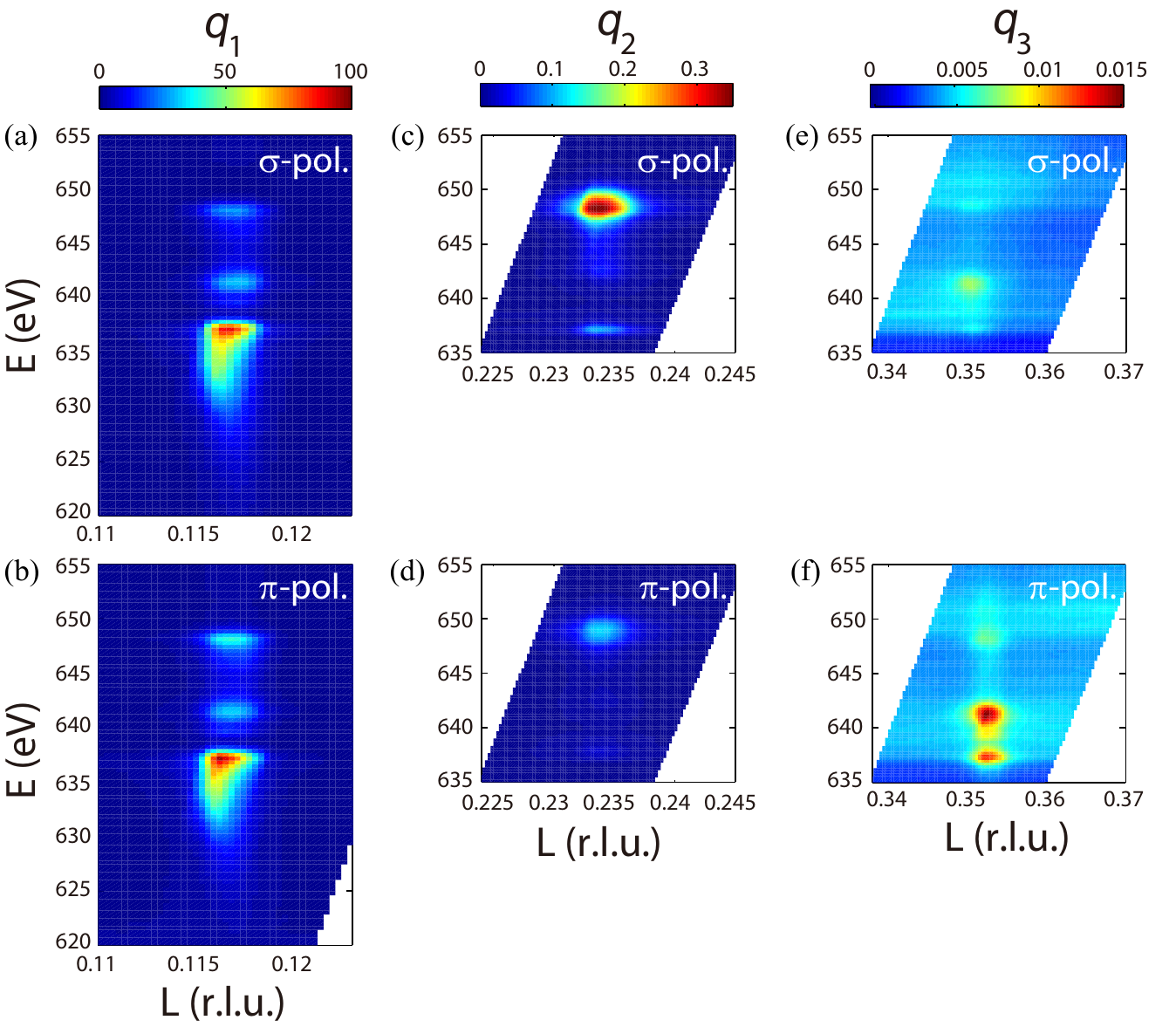}
\caption{Resonance profiles at $q_1$, $q_2$, and $q_3$ with $T$ = 20 K. The resonance profiles at each resonance position are plotted as a function of X-ray energy around the Mn $L$-edge and reciprocal lattice vector (0~0~$L$), with $\sigma$ (upper panels) or $\pi$ (lower panels) linearly polarized incident photons. (a, b) $q_1$ resonance profile. A photodiode detector was used to measure the scattering intensity $I_{\sigma}$ and $I_{\pi}$ which are comparable in intensity. (c-f) $q_2$ and $q_3$ resonance profiles measured by a channeltron detector, due to their relatively weak intensity. For $q_2$, the $I_{\sigma}$ maximum is $\sim$ 3 times of the $I_{\pi}$ maximum. For  $q_3$, the $I_{\sigma}$ maximum is about 80\% of the $I_{\pi}$ maximum. All data shown here are raw scattering intensity without background subtraction or absorption correction. The color bars indicates scattering intensity in arbitrary unit.}
\end{figure*}

To comprehensively elucidate the nature of the three resonant peaks at $q_1$, $q_2$, and $q_3$, we have plotted in Fig.~3 the resonant profiles around the three $q$ positions with $T$ = 20 K, i. e., the scattering intensities as a function of reciprocal lattice (0~0~$L$), X-ray energy, and incident photon polarization. Since there was no polarization analyzer before the detector, the polarizations of the scattered photons were not distinguished. That is, using self-explanatory  subscript convention, the detected scattered intensities for two different incident photon polarizations
$$I_{\pi}=I_{\pi\pi}+I_{\pi\sigma} \hspace{1cm}
I_{\sigma}=I_{\sigma\pi}+I_{\sigma\sigma}.$$
Clearly, the resonant profiles of $q_1$, $q_2$, and $q_3$ are very different from each other in maximal intensity, resonant energy, and polarization dependence. For example, the resonant energies for the maximum peaks at $q_1$, $q_2$, and $q_3$ are 637~eV ($\pi$ polarization), 648.2~eV ($\sigma$ polarization), and 641.2~eV ($\pi$ polarization), respectively.
 The maximal intensity at $q_3$ is $\sim$ 23 times weaker than that at $q_2$, and the maximal intensity for $q_2$ is $\sim$30 times weaker than that of $q_1$, based on  the data of the resonance peaks taken with the same detector.

\begin{figure*}[t]
\centering
\includegraphics[clip,width=12cm]{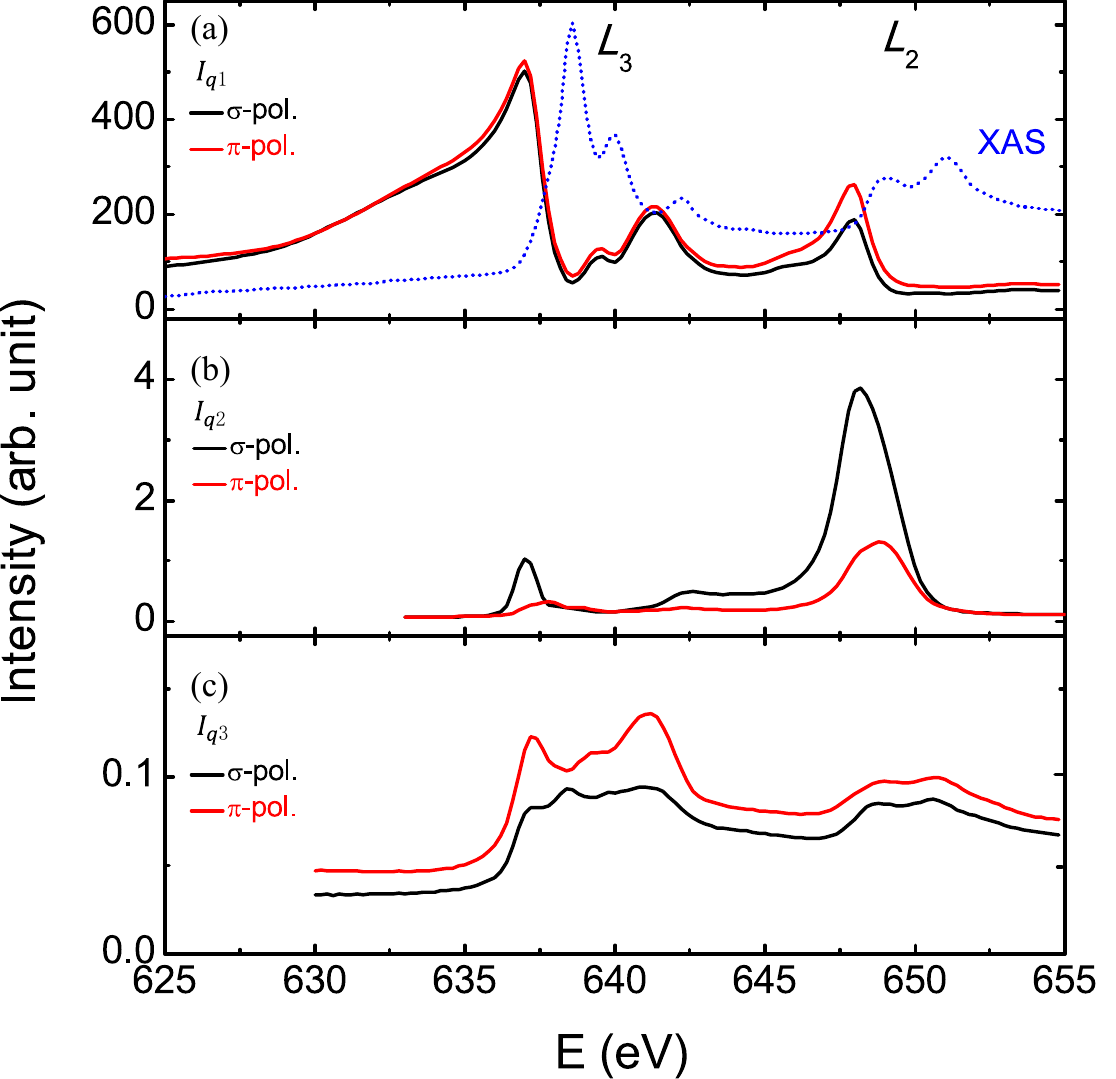}
\caption{q-integrated intensity $I_{q}$ of the three resonance peaks shown in Fig. 3. (a) $I_{q1}$ with $\sigma$ (black line) and $\pi$ (red line) polarizations. $I_{q2}$ and $I_{q3}$ are plotted in (b) and (c), respectively.}
\end{figure*}

 \begin{figure*}[]
\centering
\includegraphics[clip,width=15cm]{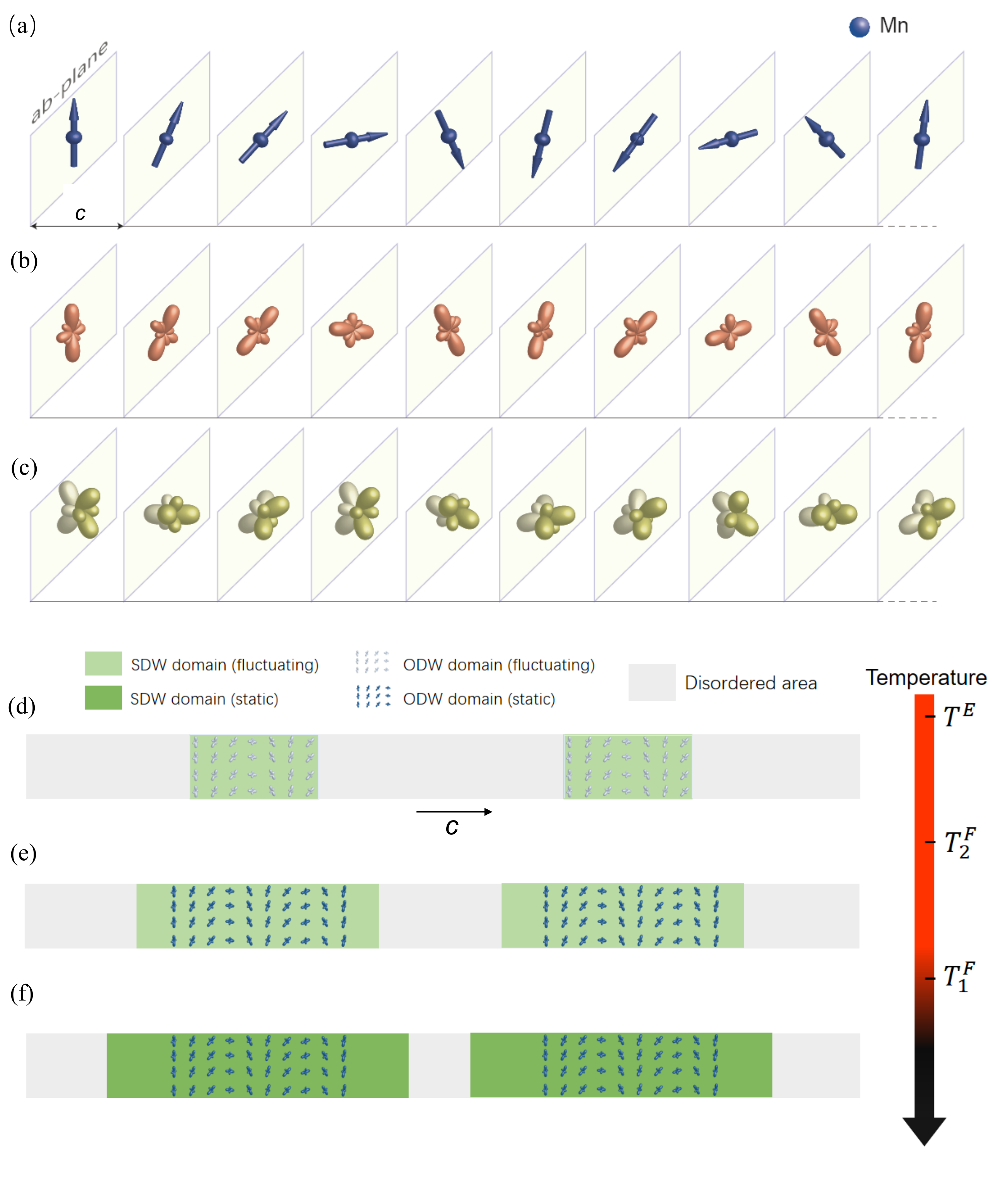}
\caption{\label{fig:epsart} An illustration of the intertwined orders  in MnP. (a) The helical spin density wave with a period of $\sim$$8.6c$ is shown with  Mn ions with equal spacing in the \textit{c} direction. (b) The orbital density wave with 1/2 the period is represented by an exaggerated mixture of $d_{xy}$ and $d_{x^2-y^2}$ orbitals that rotates together with the spin, but is symmetric under 180$^\circ$ rotation. (c) The orbital density wave with 1/3 the period is represented by an exaggerated mixture of $d_{xz}$ and $d_{yz}$ orbitals that rotates three times as fast as the spin. (d) Cartoon showing that short-ranged and fluctuating  SDW  (represented by the blue area) and ODW domains (represented by the texture) share the same region in MnP for $T^E \geq$ T$\geq T_2^F$. (e) When $T_2^F \geq$ T$\geq T_1^F$, the ODW is fully developed and stops to grow with decreased temperature, while SDW domains continue to grow. There is regions with fluctuating spin order   but without orbital order.
(f) When $ T \leq T_1^F $, the spin order is fully developed.
}
\end{figure*}

The $q$-integrated resonant profiles are compared in Fig. 4, as a function of energy and polarization.
The resonant profile of the $q_1$ peak behaves similarly in both polarizations, consistent with its origin in resonant magnetic scattering.
As illustrated by a simple analysis, magnetic scattering will result in $\pi\pi$, $\pi\sigma$  and $\sigma\pi$ scatterings with comparable intensity\cite{Jang2016,Lovesey1996}. The magnetic scattering matrix element is related to the dipole selection rules, which have the same origin as soft x-ray magnetic dichroism. This explains it having the strongest intensity amongst all three peaks.
For the $q_2$ peak,  $I_{\sigma}$ is about triple $I_{\pi}$.
Moreover, the energy positions of the $q_2$ profile differ between the $\sigma$ and $\pi$ polarizations, which is unlikely from magnetic scattering. The resonance profile of the $q_3$ peak shows moderate polarization dependence.

The saturation of the $q_1$ peak intensity  below $T_1^F$ represents the full development of the incommensurate helical spin density wave.
Remarkably,  the weak $q_2$ peak is  fully developed slightly {\slshape above} $T_1^F$, when the $q_1$ peak intensity is still two orders of magnitude smaller than its full value. It
implies that this is not a simple second-order harmonic of the spin density wave but an indication of a hidden order which is induced by magnetic fluctuations. Applying Landau theory to the phase transition and using general symmetry analysis, we extract the nature of these orders by assuming that the Landau free-energy functional depends only on the fundamental Fourier components of the two different electronic orders. The $q_1$  peak corresponds to the double helical spin density wave, $\vec S_{q_1,a}\propto (i,1,0)$ where $ a=1,2 $ denote  the two helices. As $q_2\approx 2q_1$, the hidden order, $\Pi_{q_2}$, must be coupled to a quadratic term of $\vec S_{q_1} $.  If we take $\Pi_{q_2}$ to be a scalar, a natural choice of $\Pi_{q_2}$ is the charge density wave, $\rho_{q_2}$. The lowest order coupling  in Landau theory can be written as $H_{c}=\sum_{ab}\lambda_{c,ab} \rho_{q_2}(\vec S^*_{q_1,a} \cdot \vec S^*_{q_1,b})+h.c.$.  However, such an unequal charge distribution at different Mn sites could not be found  in our LDA calculations shown in the supplementary materials. Instead, an orbital distribution with half the periodicity of  the magnetic order can be explicitly obtained in the calculation.

Here, we suggest that the hidden order must be a orbital density wave (ODW), which can  be induced by the orbital redistribution in developing the double helical spin density wave.  In a simple helically ordered state, it is well known that an orbital redistribution can be induced by spin-orbital coupling in the presence of the crystal field, and such an orbital redistribution only depends on  $|\vec S_{q_1}|$\cite{Stohr2006}, giving the X-ray magnetic linear dichroism effect (XMLD). In this case, spin-parallel and -antiparallel Mn atoms would have identical orbital (wave function) distribution, which explains why $q_2$ corresponds to an order with a period  half that of the magnetic peak. This situation resembles the stripes in cuprates, but in MnP both orders are incommensurate with respect to the lattice. We observe resonance at the Mn $2p$ to $3d$ transition, and orbital order of $3d$ electrons gives an electronic orbital order, $\boldsymbol{\Pi}_{\alpha\alpha}$. Thus a ODW is simultaneously developed in the helical spin state.

Now, in the case of {\slshape double}-helical spin order, we argue that a tiny orbital modulation can be induced by spin fluctuations even before the static helical spin order is developed. In general, the coupling between the ODW and the double-helical spin order in Landau theory can be written as

$$H_{Q}=\sum_{\alpha}\boldsymbol{\Pi}_{\alpha\alpha,q_2} [\lambda^\alpha_{Q,1} ((S^{\alpha*}_{q_1,1})^2 +(S^{\alpha*}_{q_1,2})^2) + 2\lambda^\alpha_{Q,2}  S^{\alpha*}_{q_1,1}S^{\alpha*}_{q_1,2}]
+h.c..$$

 With this coupling, there is a new phase in which $<S^{\alpha*}_{q_1,a}>=0$ but $<S^{\alpha*}_{q_1,1}S^{\alpha*}_{q_1,2}>\neq0$, which describes the locking of the magnetic fluctuations between the two helices.    In general, this phase could exist slightly above $T_s$, thus explaining the intriguing full development of the $q_2$ peak at a slightly higher temperature than the $q_1$ peak. The linear coupling between the ODW  and $S^{\alpha*}_{q_1,1}S^{\alpha*}_{q_1,a}$ must result in  $<\boldsymbol{\Pi}_{\alpha\alpha,q_2}>\neq0$ if $<S^{\alpha*}_{q_1,1}S^{\alpha*}_{q_1,2}>\neq0$.  This argument is generally known as ``order by disorder" and has been used to explain the nematicity in FeAs-based superconductors in which a similar phenomenon has been observed in the parent compounds, \textit{e.g.} BaFe$_2$As$_2$\cite{Chu2010,Yi2011}, where the nematic orbital order emerges at a slightly higher temperature than the collinear spin order due to spin fluctuations.


Since the spins lie in the $ab$ plane in the double helical phase,  the orbital moment should also be in the $ab$ plane from the coupling $H_Q$. This explains the observed polarization dependence of the $q_2$ peak and also suggests that the charge redistribution mainly occurs in the $d_{x^2-y^2}$ and $d_{xy}$ orbitals (the orbital basis in the octahedral coordination with these two orbitals in the $ab$-plane is used here for the ease of explanation, while we note Mn ions are in a tilted and distorted octahedrons made of P ions).
The observation of the weak $q_3$ peak also lends strong support to the presence of hidden ODW order since it  represents the harmonics generated by the coupling between the ODW and the helical spin order.

The intensity of the $q_3$ peak is less than 1/600 of that of the SDW   peak, yet it saturates at a slightly higher temperature. This remarkable behavior suggests that the $q_3$ peak is also non-magnetic in nature. Moreover, since  it is  slightly stronger for $\pi$-polarized than $\sigma$-polarized incident photons (Fig.~4(c)),
it should involve   $d_{xz}$, $d_{yz}$, and possibly $d_{3z^2-r^2}$ orbitals. The simplest possibility for the $q_3$ peak would the rearrangement or modulation of these orbitals induced by the coupling between the ODW ($q_2$) and the helical SDW ($q_1$) as a third-order effect.
This represents another unique ODW that is observed here for the first time.

In Figs.~5(a-c), we summarize the observed orders. The helical spin density wave is shown  in Fig.~5(a), represented by spins rotated by equally-spaced in-plane angles. In Fig.~5(b), the in-plane ODW follows the SDW with half the period, represented by a charge distribution whose axis follows the orientation of the spin. In Fig.~5(c), the out-of-plane ODW rotates with one third of the SDW period.
These helical orbital density waves in MnP  are discovered for the first time  by RSXS, and they  are intertwined with the helical spin density wave.
It is particularly noteworthy that the temperature dependencies in Fig.~2 further illustrate the intricate relation between the spin and orbital orders. The short-ranged and fluctuating orbital and spin orders share the same domain as they both emerge and grow upon lowering the temperature (Fig.5(d)). This starts from quite high temperatures,  likely related to strain distributions in the system.  Upon further cooling, as shown in Fig. 5(e), the ODW order freezes and its domains are fixed after passing its fully-developed temperature $T_2^F$. Meanwhile, the spin order is still fluctuating and its domains keep expanding, since the intensity of the spin order peak is still a few percent of its full value at low temperature. As a result,
there are regions with fluctuating spin order but no orbital order  in this temperature range.  Because the orbital order is fairly weak, as shown by the weak diffraction peak intensity, the orbital distribution can be influenced or pined by local strain or defects. Therefore,
the orbitals in these regions are disordered, and its configuration could not follow the helical spin rotation.
When temperature is further lowered below $T_1^F$, the spin order is fully developed and its domains stop to grow, as illustrated in Fig.~5(f). We note that because the out-of-plane ODW in Fig.~5(c) is the descendant of the SDW and the in-plane ODW, its domain is identical to that of the in-plane ODW.

A similar observation was made in Pr$_{0.6}$Ca$_{0.4}$MnO$_{3}$, whose spin order correlation length is longer than its charge/orbital order correlation length, which was attributed to likely decoupling between spin and charge/orbital orders\cite{Thomas2004}. The diffraction peak intensity of its orbital order is also much weaker than the spin order in  Pr$_{0.6}$Ca$_{0.4}$MnO$_{3}$, although its spin ordering temperature is lower than that of the charge/orbital order.
So the picture revealed in Figs. 5(d-f) through
our detailed temperature dependence are likely ubiquitous for systems with both spin and orbital ordering, which explains the difference in the correlation lengths in Pr$_{0.6}$Ca$_{0.4}$MnO$_{3}$.

Our findings provide an unprecedented  picture on the intricate interplay between spin and orbital orders in MnP, and show that
intertwined ordering and nematic-like ordering is ubiquitous to the phase diagrams of  unconventional superconductors, even in the case of low symmetry and incommensurate ordering.
The extraordinary  helical ODWs found here may provide a foundation for understanding the complex behaviors of spin/orbital order in  helimagnets and correlated systems in general, and for understanding the unconventional superconductivity in MnP and other related materials.

We acknowledge fruitful discussions with prof. J. L. Luo. This work is supported in part by the National Key Research and Development Program of China (Grant No.\ 2016YFA0300200 and No.\ 2017YFA0303104), the National Natural Science Foundation of China (Grant No.\ 11804137), the Science and Technology Commission of Shanghai Municipality (Grant No.\ 15ZR1402900), and the Natural Science Foundation of Shandong Province (Grant No.\ ZR2018BA026). The experiments were conducted at the REIXS beamline of the Canadian Light Source (CLS) under Proposal No.\ 24-7906, and at beamline 13-3 of Stanford Synchrotron Radiation Lightsource (SSRL); SSRL is operated by the US DOE Office of Basic Energy Science.

$^\dag$ jphu@iphy.ac.cn

$^\ast$ dlfeng@fudan.edu.cn






\begin{figure*}[t]
\centering
\includegraphics[clip,width=19cm]{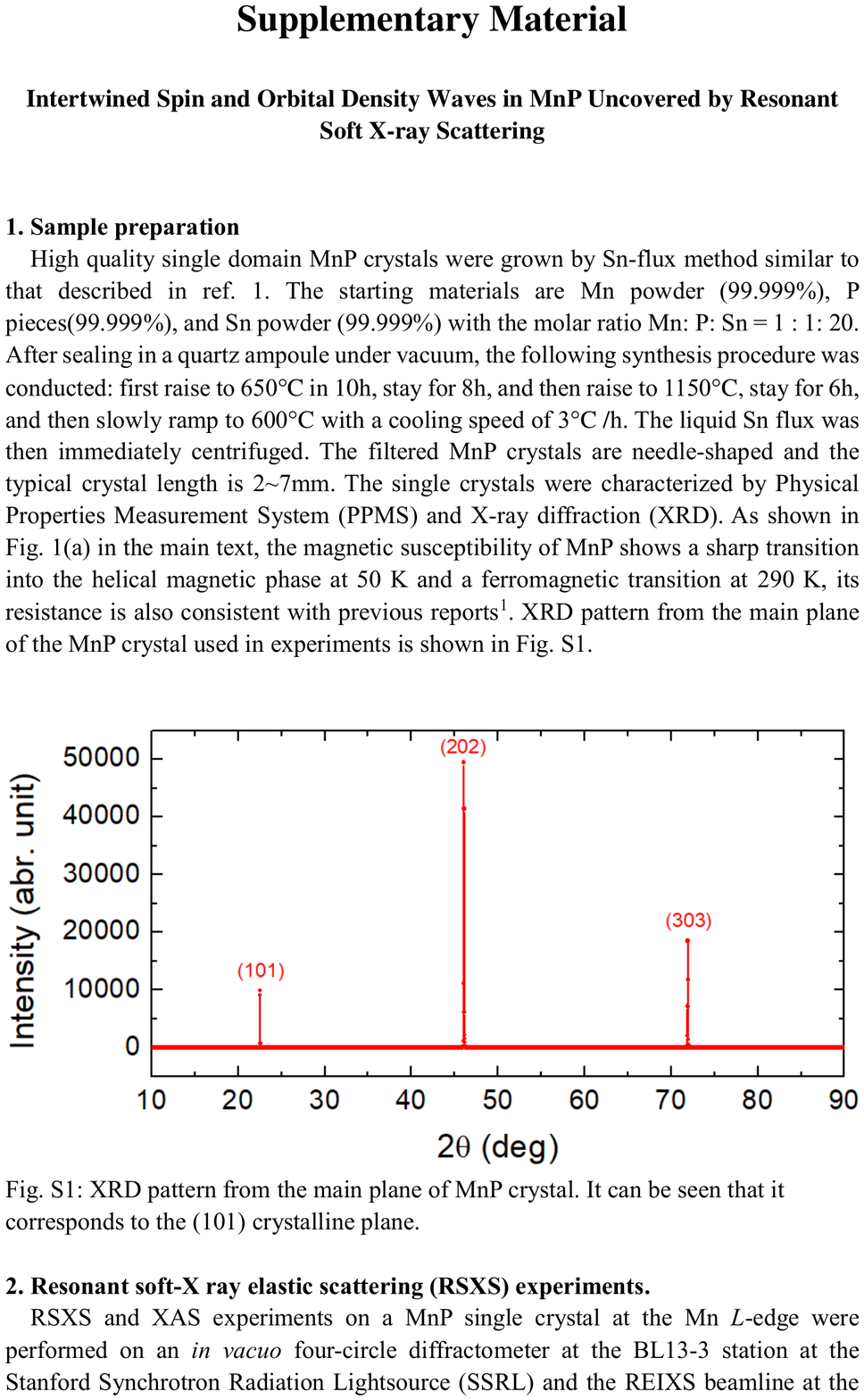}
\caption{}
\end{figure*}

\begin{figure*}[t]
\centering
\includegraphics[clip,width=19cm]{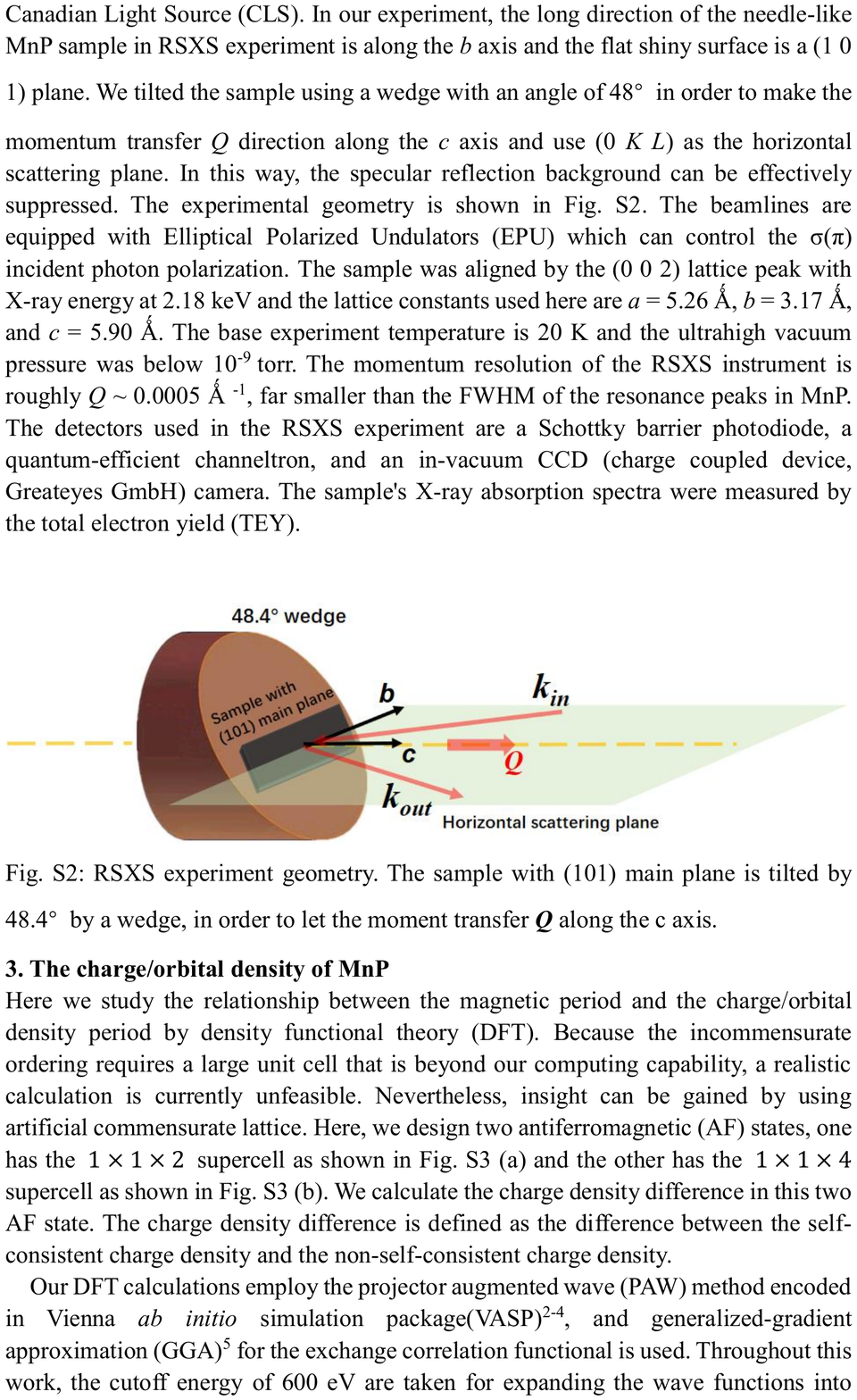}
\caption{}
\end{figure*}

\begin{figure*}[t]
\centering
\includegraphics[clip,width=19cm]{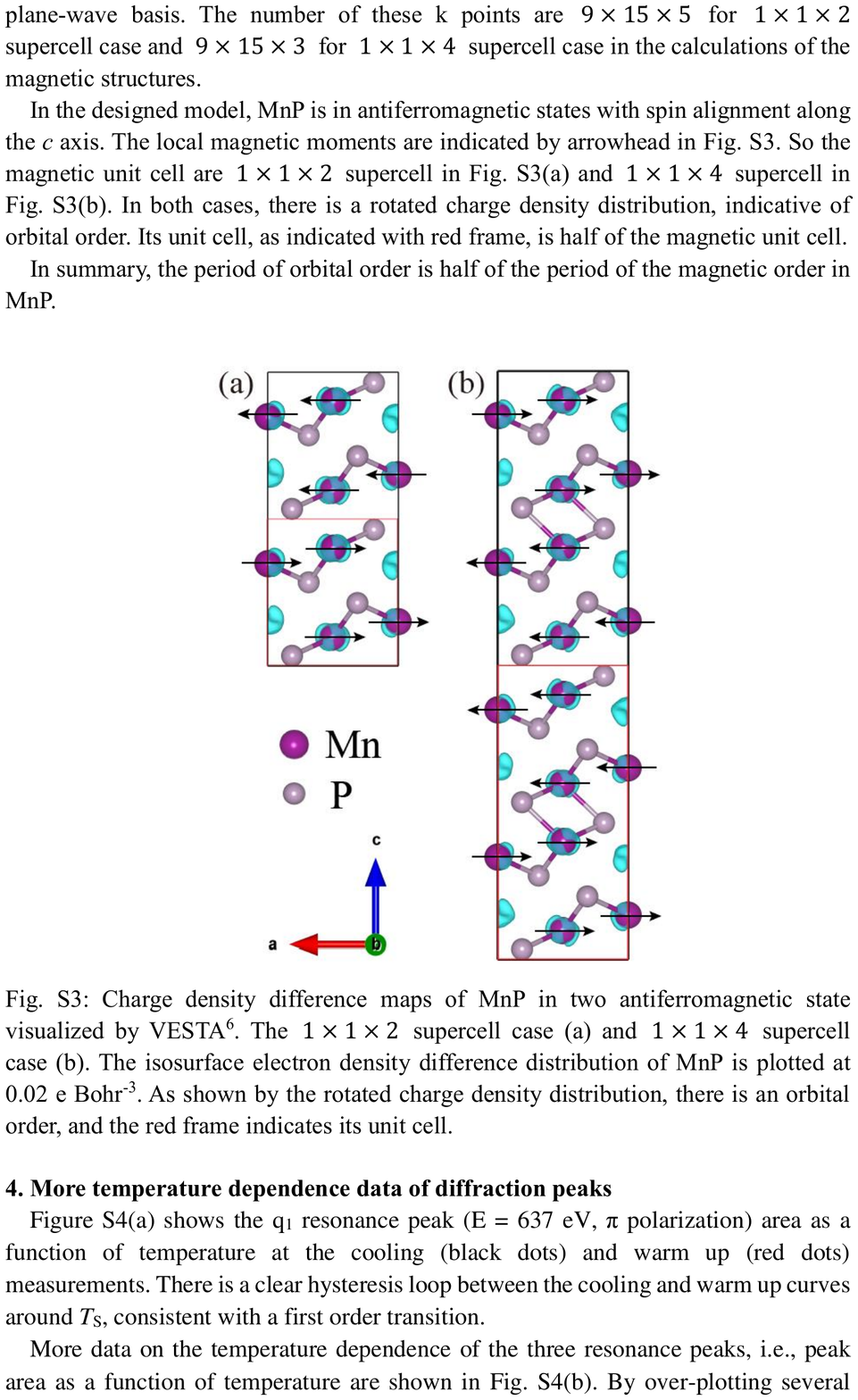}
\caption{}
\end{figure*}

\begin{figure*}[t]
\centering
\includegraphics[clip,width=19cm]{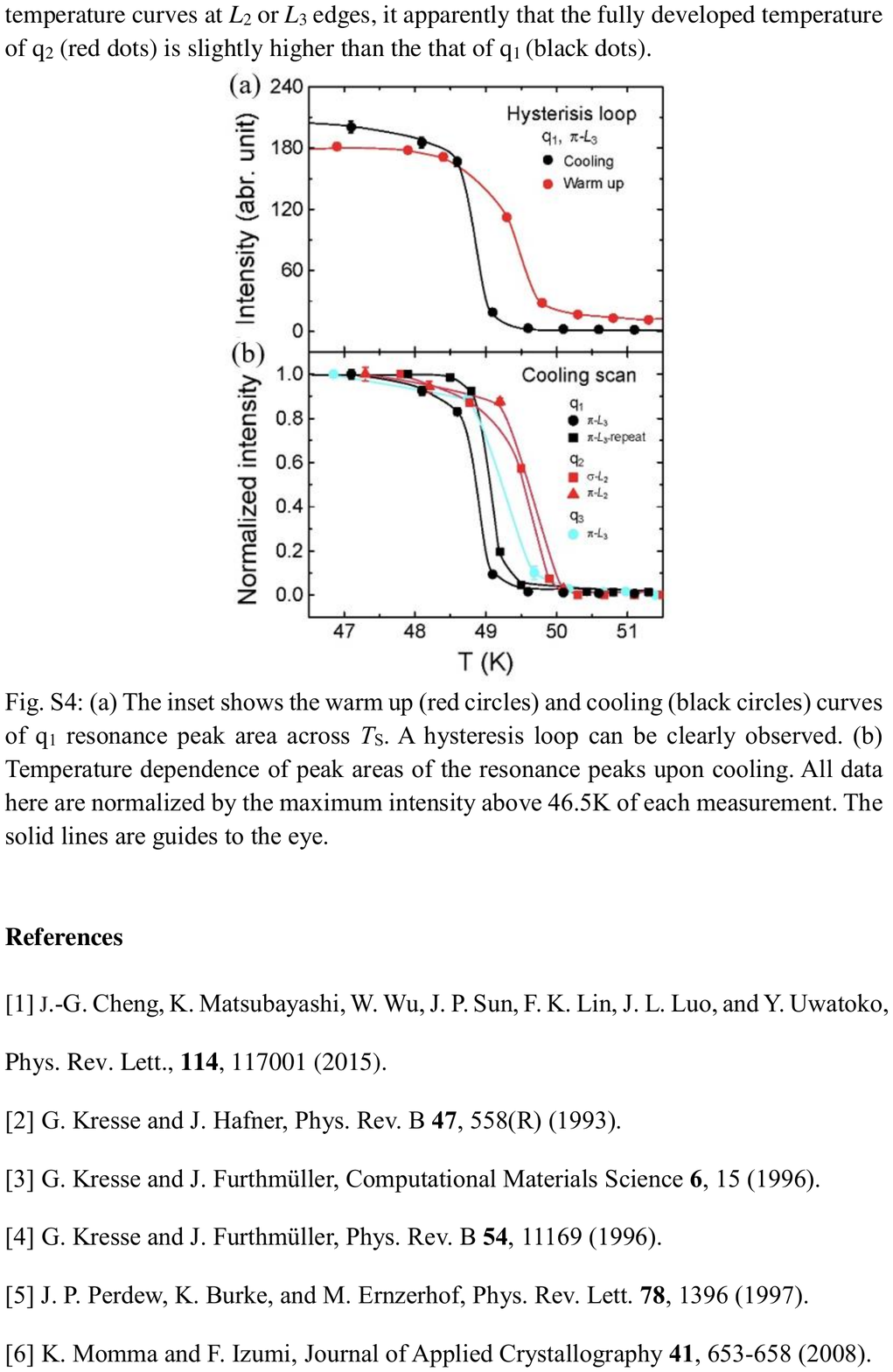}
\caption{}
\end{figure*}

\end{document}